\documentclass[conference]{IEEEtran}
\IEEEoverridecommandlockouts

\usepackage{cite}
\usepackage{amsmath,amssymb,amsfonts,lipsum}
\usepackage{graphicx}
\usepackage{textcomp}
\usepackage{xcolor}
\usepackage{caption}
\usepackage{subcaption}
\usepackage{bbm}
\usepackage{graphicx}
\usepackage{booktabs}
\usepackage{enumitem}
\usepackage{multicol}
\usepackage{algpseudocode}
\usepackage{booktabs}
\usepackage{multirow}
\usepackage[hyphens]{url}
\usepackage[hidelinks]{hyperref}
\hypersetup{breaklinks=true}
\usepackage{mathrsfs}

\newtheorem{theorem}{Theorem}[section]

\newtheorem{remark}{Remark}
\usepackage{algorithm, algpseudocode}

\newcommand{\scrF}{\mathscr{F}}

\def\BibTeX{{\rm B\kern-.05em{\sc i\kern-.025em b}\kern-.08em
    T\kern-.1667em\lower.7ex\hbox{E}\kern-.125emX}}
\begin{document}

\title{Comparing Bitcoin and Ethereum tail behavior via Q–Q analysis of cryptocurrency returns}
\author{\IEEEauthorblockN{1\textsuperscript{st} A. H. Nzokem}
hilaire77@gmail.com}

\maketitle
\begin{abstract}
The cryptocurrency market presents both significant investment opportunities and higher risks relative to traditional financial assets. This study examines the tail behavior of daily returns for two leading cryptocurrencies, Bitcoin and Ethereum, using seven-parameter estimates from prior research \cite{jrfm17120531}, which applied the Generalized Tempered Stable (GTS) distribution. Quantile-quantile (Q–Q) plots against the Normal distribution reveal that both assets exhibit heavy-tailed return distributions. However, Ethereum consistently shows a greater frequency of extreme values than would be expected under its Bitcoin-modeled counterpart, indicating more pronounced tail risk.
\end{abstract}
\begin{IEEEkeywords}
Cryptocurrency returns; Tail risk; Bitcoin; Ethereum; Generalized Tempered Stable (GTS) distribution; Quantile–Quantile (Q–Q) plots; Heavy tails  
\end{IEEEkeywords}

\section{Introduction}
\noindent
Cryptocurrencies, which are digital assets secured through cryptographic techniques \cite{lewis2018basics}, have reshaped the financial landscape.  The first cryptocurrency, Bitcoin, was introduced in 2009 by Satoshi Nakamoto. The idea behind Bitcoin was to create a peer-to-peer electronic payment system that allows online payments to be sent directly from one party to another without going through a financial institution \cite{nakamoto2008bitcoin}. \\
Ethereum, on the other hand, was introduced in 2015 and emerged as the first decentralized platform enabling programmable smart contracts \cite{Najafov2022}. Importantly, Ethereum is not directly comparable to Bitcoin. While Bitcoin is primarily seen as an alternative currency and falls under the Currencies Crypto Sector, Ethereum serves as the cornerstone of the Smart Contract Platforms Crypto Sector, enabling decentralized applications (dApps) and supporting decentralized finance (DeFi) protocols and Non-Fungible Tokens (NFTs) through its native Ether token, often referred simply as Ethereum\cite{grayscale_eth_og}.\\
According to CoinMarketCap, as of mid‑2025, the total cryptocurrency market capitalization stands at around $\$3.31$ trillion, with Bitcoin (BTC) holding a market cap of $\$2.06$ trillion, making up $62.11\%$ of the total market. In comparison, Ethereum has a market capitalization of approximately $\$292$ billion, representing $7.89\%$ of the overall market\cite{coinmarketcap2025}.\\

The growing interest in cryptocurrencies from both academic research and the business sector has increased efforts to analyse the financial risk profiles of the daily returns of Bitcoin and Ethereum. Our analysis goes beyond traditional metrics such as central tendency and dispersion, aiming to uncover more nuanced distributional characteristics of their high-frequency returns.\\

In this study, the seven-parameter Generalized Tempered Stable (GTS) distribution is used as the underlying distribution for the daily returns of Bitcoin and Ethereum. The methodology for fitting the GTS distribution to financial data was outlined in a previous study \cite{jrfm17120531}, which also provided statistical estimates for the seven parameters in the context of S\&P 500, SPY ETF, Bitcoin, and Ethereum. The goodness-of-fit for these estimations was assessed using Kolmogorov-Smirnov, Anderson-Darling, and Pearson's chi-squared tests, with results demonstrating a significant fit and high p-values. Moreover, the GTS distribution outperformed other models such as the Kobol, Carr-Geman-Madan-Yor, and Bilateral Gamma distributions \cite{jrfm17120531}. Further details on the GTS distribution fitting process can be found in \cite{jrfm17120531, mca29030044, nzokem2024self}. While the Generalized Hyperbolic (GH) distribution and its subclasses \cite{jrfm10020012, ZHANG2019120900, Najafov2022} are also frequently used for modeling cryptocurrencies, the GTS distribution stands out due to its flexibility and ability to capture the heavy-tailed nature of financial data, as opposed to the GH distribution which has a closed-form expression for the probability density function.\\

The paper is structured as follows: Section 2 provides an overview of the GTS distribution and its seven-parameter estimation results.  And Section 3 discusses the results from the Quantile-Quantile (Q-Q) plot analysis.

\section{Methodology and Parameter Estimation Results}
\noindent
The L\'evy measure of the GTS distribution ($V(dx)$) is defined in (\ref{eq:l2}) as a product of a tempering function $q(x)$ and a L\'evy measure of the $\alpha$-stable distribution $V_{stable}(dx)$:
\begin{equation}
 \begin{aligned}
q(x) = e^{-\lambda_{+}x} \boldsymbol{1}_{x>0} + e^{-\lambda_{-}|x|} \boldsymbol{1}_{x<0} \\
V_{stable}(dx) =\left(\alpha_{+}\frac{1}{x^{1+\beta_{+}}} \boldsymbol{1}_{x>0} + \alpha_{-}\frac{1}{|x|^{1+\beta_{-}}} \boldsymbol{1}_{x<0} \right) dx \\ \label{eq:l2}
V(dx)=\left(\alpha_{+}\frac{e^{-\lambda_{+}x}}{x^{1+\beta_{+}}} \boldsymbol{1}_{x>0} + \alpha_{-}\frac{e^{-\lambda_{-}|x|}}{|x|^{1+\beta_{-}}} \boldsymbol{1}_{x<0}\right) dx.
 \end{aligned}
 \end{equation}

\noindent
where $0\leq \beta_{+}\leq 1$, $0\leq \beta_{-}\leq 1$, $\alpha_{+}\geq 0$, $\alpha_{-}\geq 0$, $\lambda_{+}\geq 0$ and  $\lambda_{-}\geq 0$.\\
The six parameters hold significant interpretations. $\beta_{+}$ and $\beta_{-}$ are the indexes of stability bounded below by zero and above by 2 \cite{borak2005stable}. They capture the peakedness of the distribution similarly to the $\beta$-stable distribution. If $\beta$ increases (decreases), then the peakedness decreases (increases). $\alpha_{+}$ and $\alpha_{-}$ are the scale parameters, also called the process intensity \cite{boyarchenko2002non}; they determine the arrival rate of jumps for a given size. $\lambda_{+}$ and $\lambda_{-}$ control the decay rate on the positive and negative tails. Additionally, $\lambda_{+}$ and $\lambda_{-}$ are also skewness parameters. If $\lambda_{+}>\lambda_{-}$ ($\lambda_{+}<\lambda_{-}$), then the distribution is skewed to the left (right), and if $\lambda_{+}=\lambda_{-}$, then it is symmetric \cite{rachev2011financial, fallahgoul2019quantile}.\\
For more details on the tempering function and the associated  L\'evy measure, refer to thr following works \cite{kuchler2013tempered, rachev2011financial, bianchi2019handbook}.\\

\noindent
The activity process of the GTS distribution can be studied from the integral (\ref{eq:l3}) of the L\'evy measure (\ref{eq:l2}):
\begin{equation}
\begin{aligned}
 \int_{-\infty}^{+\infty} V(dx) =+\infty \quad \text{if }{0\leq \beta_{+} < 1 \wedge 0\leq\beta_{-} <1} . \label{eq:l3}
 \end{aligned}
 \end{equation}

\noindent
As shown in Eq (\ref{eq:l3}), if $0\leq \beta_{+}<1$ and $0\leq \beta_{-}< 1$, the L\'evy density ($V(dx)$) is not integrable as it goes off to infinity too rapidly as $x$ goes to zero, due to a large number of very small jumps. The GTS distribution is said to be an infinite activity process with infinite jumps in any given time interval. \\

\noindent
In addition to the infinite activities process,  the variation of the process can be studied by solving the following integral\cite{kyprianou2014fluctuation}:
\begin{equation}
\begin{aligned}
  \int_{-1}^{1} |x|V(dx) <+\infty  \quad \text{if }{0\leq \beta_{+} < 1 \wedge 0\leq\beta_{-} <1} . \label{eq:l42}
 \end{aligned}
 \end{equation}
Refer to \cite{nzokem2024self} for further development Eq (\ref{eq:l42}) \\

\noindent
As shown in Eq (\ref{eq:l42}), $GTS (\beta_{+}, \beta_{-}, \alpha_{+}, \alpha_{-}, \lambda_{+}, \lambda_{-})$ is a finite variation process, and generates a type B L\'evy process \cite{barndorff2001levy}, which is a purely non-Gaussian infinite activity L\'evy process of finite variation whose sample paths have an infinite number of small jumps and a finite number of large jumps in any finite time interval. In particular, being of bounded variation shows that $GTS (\beta_{+}, \beta_{-}, \alpha_{+}, \alpha_{-}, \lambda_{+}, \lambda_{-})$ can be written as the difference of two independent subordinators \cite{kyprianou2014fluctuation, tankov2010financial}.
\begin{equation*}
 \begin{aligned}
X= X_{+}  -  X_{-} . \label {eq:l55}
  \end{aligned}
\end{equation*}
where $X_{+} \sim TS(\beta_{+}, \alpha_{+},\lambda_{+})$ and $X_{-} \sim TS(\beta_{-}, \alpha_{-},\lambda_{-})$ are subordinators.\\
\noindent
By  adding a drift parameter, we have the following expression
 \begin{align}
Y=\mu + X \sim GTS(\mu, \beta_{+}, \beta_{-}, \alpha_{+}, \alpha_{-}, \lambda_{+}, \lambda_{-}) . \label {eq:l5}
  \end{align}

\begin{theorem}\label{lem5} \ \\
Consider a variable $Y \sim GTS(\mu, \beta_{+}, \beta_{-}, \alpha_{+},\alpha_{-}, \lambda_{+}, \lambda_{-})$. The characteristic exponent can be written as:
 \begin{equation}
\begin{aligned}
\psi(\xi)&=\mu\xi i+\alpha_{+}\Gamma(-\beta_{+})\left((\lambda_{+} - i\xi)^{\beta_{+}}-{\lambda_{+}}^{\beta_{+}}\right)+\\
&\alpha_{-}\Gamma(-\beta_{-})\left((\lambda_{-}+i\xi)^{\beta_{-}}-{\lambda_{-}}^{\beta_{-}}\right). \label{eq:l6}
 \end{aligned}
 \end{equation}
\end{theorem}

See \cite{mca29030044, jrfm17120531, nzokem2021fitting} for Theorem \ref{lem5} proof.

\subsection{Parameter Estimation Results: Bitcoin versus Ethereum}
\noindent
The GTS parameter estimation results for Bitcoin and Ethereum are summarized in Tables \ref{tab1} and \ref{tab2}, respectively. The brackets indicate the asymptotic standard errors, calculated using the inverse of the Hessian matrix.. Both Bitcoin and Ethereum's log-likelihood estimates show that  $\mu$ is negative, but the asymptotic standard error is quite large, suggesting that $\mu$ is not statistically significant at 5\% level.

 \begin{table}[ht]
\centering
\caption{GTS Parameter Estimation for Bitcoin}
\label{tab1}
\begin{tabular}{@{} c|c|c|c|c @{}}
\toprule
\multicolumn{1}{c|}{\textbf{Model}} & \multicolumn{1}{c|}{\textbf{Parameter}} & \multicolumn{1}{c|}{\textbf{Estimate}} & \multicolumn{1}{c|}{\textbf{Std Err}} &  \multicolumn{1}{c}{\textbf{$Pr(Z >|z|)$}}   \\ \toprule
\multirow{10}{*}{\textbf{GTS}} & \multirow{1}{*}{\textbf{$\mu$}} &\multirow{1}{*}{-0.121571} & \multirow{1}{*}{(0.375)} & \multirow{1}{*}{$7.5*10^{-01}$}  \\
 & \multirow{1}{*}{\textbf{$\beta_{+}$}} & \multirow{1}{*}{0.315548} & \multirow{1}{*}{(0.136)}  & \multirow{1}{*}{$2.0*10^{-02}$} \\

 &\multirow{1}{*}{ \textbf{$\beta_{-}$}} & \multirow{1}{*}{0.406563} & \multirow{1}{*}{(0.117)} & \multirow{1}{*}{$4.9*10^{-04}$}  \\

 & \multirow{1}{*}{\textbf{$\alpha_{+}$}} & \multirow{1}{*}{0.747714} & \multirow{1}{*}{(0.047)} & \multirow{1}{*}{$6.2*10^{-56}$}   \\

 & \multirow{1}{*}{\textbf{$\alpha_{-}$}} & \multirow{1}{*}{0.544565} & \multirow{1}{*}{(0.037)} & \multirow{1}{*}{$4.8*10^{-48}$}  \\

 &\multirow{1}{*}{\textbf{$\lambda_{+}$}} & \multirow{1}{*}{0.246530} & \multirow{1}{*}{(0.036)} & \multirow{1}{*}{$4.9*10^{-12}$} \\

 & \multirow{1}{*}{\textbf{$\lambda_{-}$}} & \multirow{1}{*}{0.174772} & \multirow{1}{*}{(0.026)} & \multirow{1}{*}{$2.2*10^{-11}$}  \\ \bottomrule
\end{tabular}
\end{table}
\noindent
For Bitcoin returns, the skewness parameters ($\lambda_{+}$, $\lambda_{-}$), the process intensity parameters ($\alpha_{+}$, $\alpha_{-}$), the index of stability parameters ($\beta_{+}$, $\beta_{-}$) are all statistically significant at the 5\% level.\\

 \noindent
 Table \ref{tab2} provides a summary of the estimation results for Ethereum returns. The parameters for Ethereum returns data are statistically significant at 5\%, except $\mu$ and $\beta_{-}$. 

\begin{table}[ht]
\centering
\caption{GTS Parameter Estimation for Ethereum}
\label{tab2}
\begin{tabular}{@{} c|c|c|c|c@{}}
\toprule
\multicolumn{1}{c|}{\textbf{Model}} & \multicolumn{1}{c|}{\textbf{Param}} & \multicolumn{1}{c|}{\textbf{Estimate}} & \multicolumn{1}{c|}{\textbf{Std Err}} & \multicolumn{1}{c}{\textbf{$Pr(Z >|z|)$}} \\ \toprule

\multirow{10}{*}{\textbf{GTS}} & \multirow{1}{*}{\textbf{$\mu$}} &  \multirow{1}{*}{-0.4854} & \multirow{1}{*}{(1.008)} & \multirow{1}{*}{$6.3*10^{-01}$} \\

 & \multirow{1}{*}{\textbf{$\beta_{+}$}} & \multirow{1}{*}{0.3904} & \multirow{1}{*}{(0.164)} & \multirow{1}{*}{$1.7*10^{-02}$} \\

 &\multirow{1}{*}{ \textbf{$\beta_{-}$}} & \multirow{1}{*}{0.4045} & \multirow{1}{*}{(0.210)} & \multirow{1}{*}{$5.4*10^{-02}$} \\

 & \multirow{1}{*}{\textbf{$\alpha_{+}$}} & \multirow{1}{*}{0.9582} & \multirow{1}{*}{(0.106)}  & \multirow{1}{*}{$1.1*10^{-19}$} \\

 & \multirow{1}{*}{\textbf{$\alpha_{-}$}} & \multirow{1}{*}{0.8005} & \multirow{1}{*}{(0.110)} & \multirow{1}{*}{$4.2*10^{-13}$}\\

 &\multirow{1}{*}{\textbf{$\lambda_{+}$}} & \multirow{1}{*}{0.1667} & \multirow{1}{*}{(0.029)} &\multirow{1}{*}{$1.1*10^{-08}$} \\

 & \multirow{1}{*}{\textbf{$\lambda_{-}$}} & \multirow{1}{*}{0.1708} & \multirow{1}{*}{(0.036)} & \multirow{1}{*}{$2.5*10^{-06}$} \\ \bottomrule
\end{tabular}%
\end{table}

However, the difference ($\lambda_{+} - \lambda_{-}$), the difference ($\alpha_{+} - \alpha_{-}$) are not statistically significant. Contrary to the Bitcoin return, the Ethereum return has a larger process intensity, which provides evidence that Ethereum has a lower level of peakedness and a higher level of thickness. \\

\begin{remark}
The methodology for fitting a seven-parameter GTS distribution and the parameter estimation results can be further studied in \cite{jrfm17120531}, notably
\begin{itemize}
  \item Bitcoin (BTC) and Ethereum (ETH) prices were extracted from CoinMarketCap. The period spans from April 28, 2013, to July 04, 2024, for Bitcoin, and from August 07, 2015, to July 04, 2024, for Ethereum.
    \item  The log-likelihood, Akaike's information Criteria (AIC), and Bayesian information criteria (BIK) statistics show that the GTS distribution with seven parameters performs better than the two-parameter Normal distribution (GBM)
  \item The goodness-of-fit to GTS distribution was assessed through Kolmogorov-Smirnov, Anderson-Darling, and Pearson's chi-squared tests. 
   \item It was shown that the GTS distribution outperforms the Kobol distributionKobol ($\beta=\beta_{+}=\beta_{-}$), the Carr–Geman–Madan–Yor (CGMY) distribution ($\lambda=\lambda_{+} =\lambda_{-}$ and $\beta=\beta_{+} = \beta_{-}$), and Bilateral Gamma distribution ($\beta_{+}=\beta_{-}=0$).
\end{itemize}
\end{remark}
\section{Quantile-Quantile (Q-Q) plot Analysis}
 The GTS distribution lacks a closed-form probability density function (\ref{eq:l821}), making direct sampling challenging. However, we know the closed form of the Fourier of the density function, $\scrF[f]$ (\ref{eq:l6}), and the relationship in (\ref{eq:l822}) provides the Fourier of the cumulative distribution function, $\scrF[F]$. And the GTS distribution function, $F(x)$, was computed from the inverse of the Fourier of the cumulative distribution, $\scrF[F]$, in (\ref{eq:l823}):
  \begin{align}
 Y &\sim GTS(\textbf{$\mu$}, \textbf{$\beta_{+}$}, \textbf{$\beta_{-}$}, \textbf{$\alpha_{+}$},\textbf{$\alpha_{-}$}, \textbf{$\lambda_{+}$}, f\textbf{$\lambda_{-}$}) \label{eq:l820}\\
   F(x)&= \int_{-\infty}^{x} f(t) \mathrm{d}t \hspace{5mm}  \hbox{$f$ is the density function of $Y$} \label{eq:l821}\\
 \scrF[F](x)&=\frac{\scrF[f](x)}{ix} + \pi\scrF[f](0)\delta (x) \label{eq:l822}\\
F(x) &= \frac{1} {2\pi}\int_{-\infty}^{+\infty}\! \frac{\scrF[f](y)}{iy}e^{ixy}\, \mathrm{d}y + \frac{1}{2}\label{eq:l823}
 \end{align}
See Appendix A in \cite{nzokem2021fitting} for (\ref{eq:l823}) proof.\\

  \begin{theorem} \label{lem08} \ \\
{Let} 
 a probability cumulative function $F(x)$ be at least four times continuously differentiable and let $\left(F_{j}\right)_{1 \leq j \leq m}$ be a sample of $F(x)$  on a sequence of evenly spaced input values $\left(x_{j}\right)_{1 \leq j \leq m}$, with  $F(x_{j})=F_{j}$.  We also consider $x_{\alpha}$, a $\alpha^{th}$ quantile defined by $F(x_{\alpha})=\alpha$, with $x_{i}<x_{\alpha}<x_{i+1}$ and $F_{i}<F(x_{\alpha})<F_{i+1}$. {There } exists a unique value, $y \in (0,1)$, and $b_{0}$, $b_{1}$, $b_{2}$, $b_{3}$, $b_{4}$ coefficients such that $y$ is a solution of the polynomial equation of degree 4 (\ref{eq:l85}).
  \begin{equation}
b_{0} + b_{1} y  + b_{2} y^{2} + b_{3} y^{3} + b_{4} y^{4} = 0
\label{eq:l85}
\end{equation}

The $\alpha^{th}$ quantile ($x_{\alpha}$) can be written as follows:
\begin{equation}
 x_{\alpha}=x_{i} + y(x_{i+1} - x_{i}).
\label{eq:l86}
\end{equation}
\end{theorem}
Refer to \cite{mca29030044} for the proof of Theorem \ref{lem08} 

\begin{figure}[h]
\vspace{-0.3cm}
\captionsetup{justification=centering}
\hspace{-0.5cm}
\centering 
 \begin{subfigure}[b]{0.7\linewidth}
    \includegraphics[width=\linewidth]{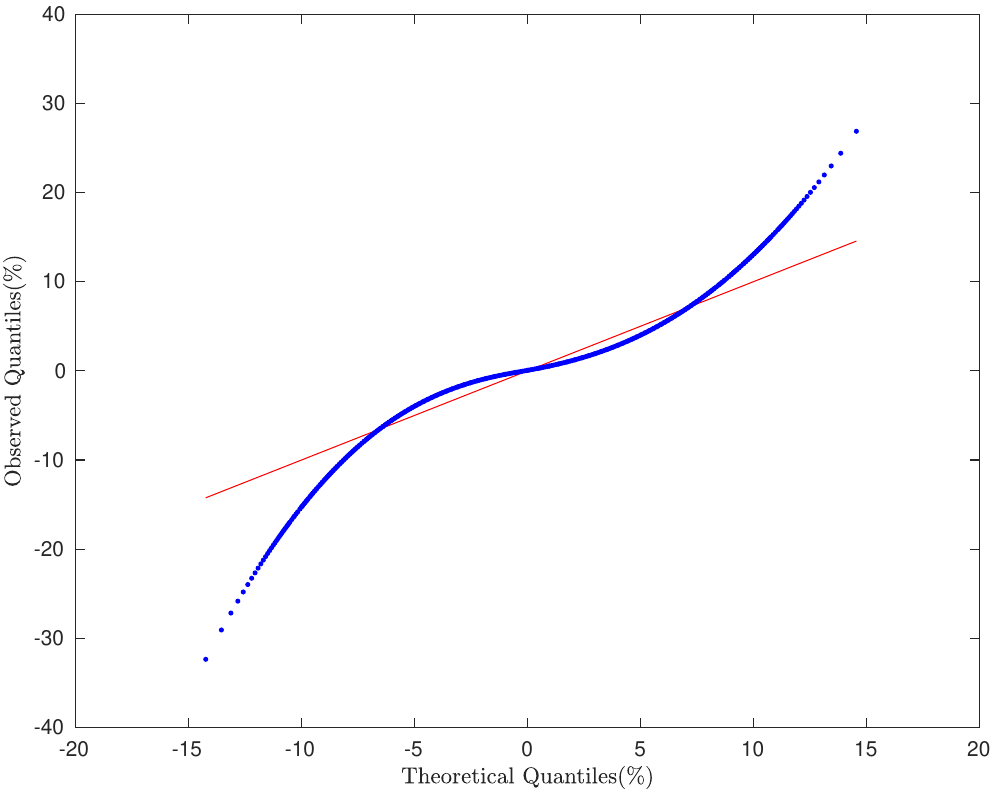}
\vspace{-0.5cm}
     \caption{heavly tailed Bitcoin}
         \label{fig:181}
  \end{subfigure}
  \begin{subfigure}[b]{0.7\linewidth}
    \includegraphics[width=\linewidth]{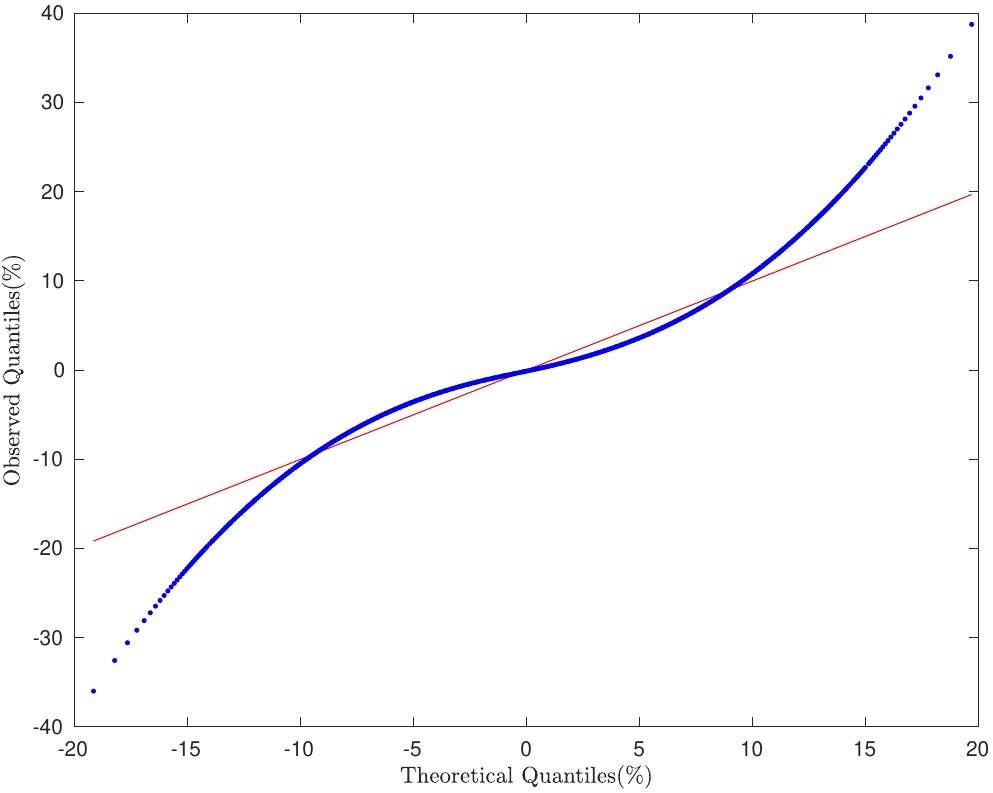}
\vspace{-0.5cm}
     \caption{heavily tailed Ethereum}
         \label{fig:182}
          \end{subfigure}
\vspace{-0.5cm}
  \caption{Quantile-Quantile (Q-Q) plots}  
  \label{fig:18}
\vspace{-0.5cm}
\end{figure}

\begin{figure}[h]
\vspace{-0.3cm}
\captionsetup{justification=centering}
\hspace{-0.3cm}
\centering 
  \begin{subfigure}[b]{0.7\linewidth}
    \includegraphics[width=\linewidth]{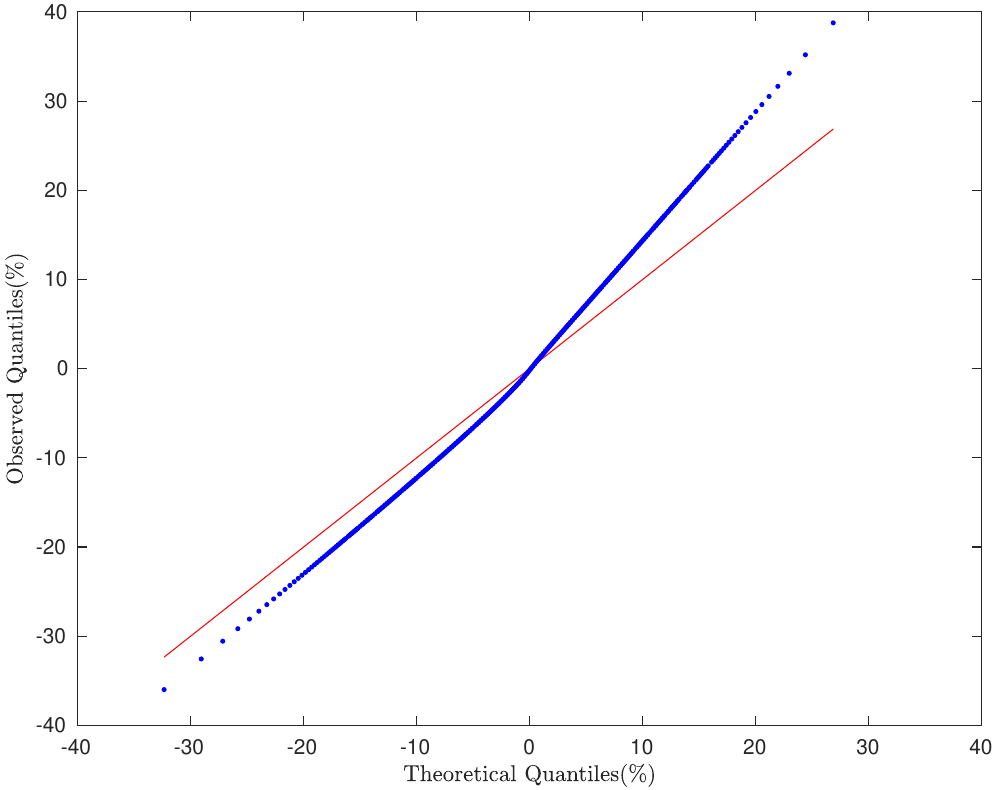}
\vspace{-0.5cm}
     \caption{heavily tailed Ethereum}
         \label{fig:191}
          \end{subfigure}
 \begin{subfigure}[b]{0.7\linewidth}
    \includegraphics[width=\linewidth]{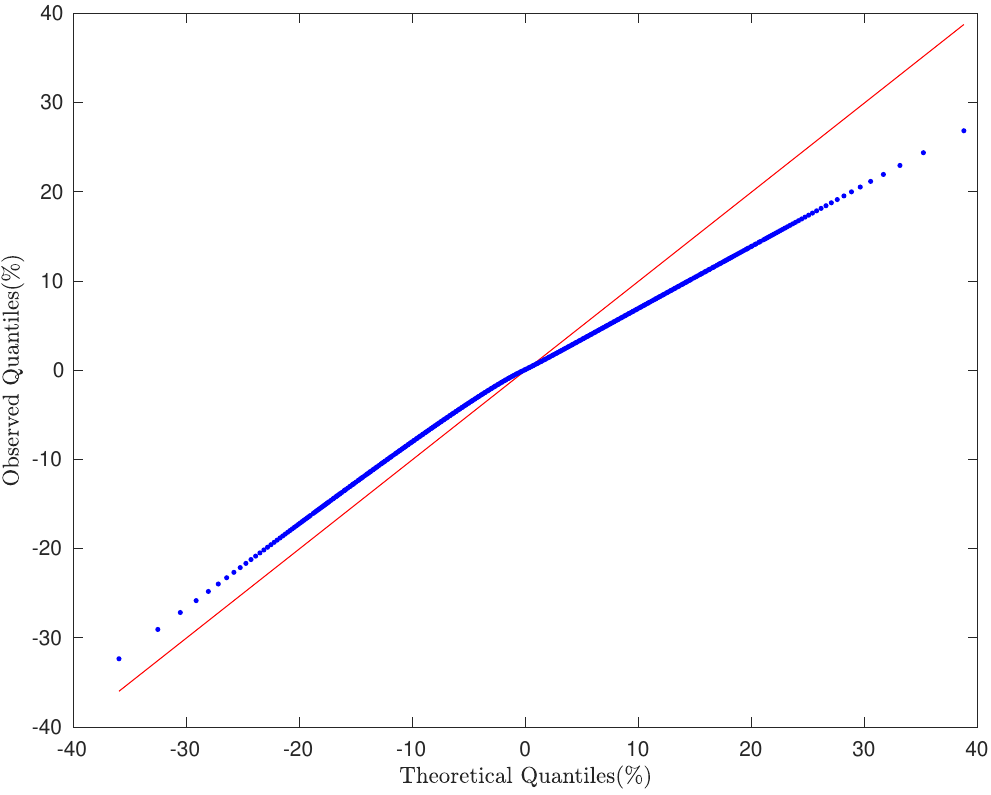}
\vspace{-0.5cm}
     \caption{lighly tailed Bitcoin}
         \label{fig:192}
          \end{subfigure}
\vspace{-0.3cm}
  \caption{Quantile-Quantile (Q-Q) plot}  
  \label{fig:19}
\vspace{-0.5cm}
\end{figure}

Panels (~ \ref{fig:191} ) and ( Panel (~\ref{fig:192}) ) compare the observed distributions of Bitcoin and Ethereum returns against a theoretical Normal distribution using Q–Q plots. Two deviations from normality stand out:
\begin{itemize}
  \item  \textbf{$S$-shaped curvature in the central region:} This “S” pattern typically occurs with asymmetric distributions
  \item \textbf{Tail-specific deviations at the ends:} The plotted points in the blue tails diverge from the red reference line: The lower tail turns downward while the upper tail curves upward. Such behavior is an indicator of heavy-tailed characteristics.\\
\end{itemize}
Together, these features suggest that both Bitcoin and Ethereum exhibit more extreme values than predicted by a Normal distribution. Therefore, Bitcoin and Ethereum are heavy-tailed distributions.\\

\begin{remark}
The theoretical quantiles ( represented by the red line in the Q-Q plot) and the observed quantiles were computed by the Enhanced Fast Fractional Fourier Transform (FRFT) scheme. The Enhanced Fast FRFT scheme improves the accuracy of the one-dimensional fast FRFT by leveraging closed Newton-Cotes quadrature rules\cite{Nzokem_2021, aubain2020}. For more details on the methodology and its applications, refer to \cite{nzokem2023enhanced, nzokem2023european, Nzokem_Montshiwa_2023, nzokem2024a, nzokem2022bis, nzokem_2022}.\\
 \end{remark}
Fig~\ref{fig:19} used the Bitcoin distribution in Panel (\ref{fig:191}) and the Ethereum distribution in Panel (\ref{fig:192}) as a theoretical distribution. 

As shown in Fig~\ref{fig:19},  both Panels appear distinct but convey identical insights about tail heaviness. Ethereum’s Q–Q plot displays more extreme values in both tails than Bitcoin’s, indicating that Ethereum has a heavier-tailed distribution. In other words, compared to Ethereum, Bitcoin exhibits lighter tails, therefore, less propensity for extreme outcomes.\\
\newpage
  \begin{remark}
In a quantile-quantile (Q-Q) plot, the quantiles of an observed distribution are plotted against the quantiles of the theoretical distribution. If the two distributions are similar, then the quantiles will also be similar and the points will fall close to the line $X =Y$. Any deviation from the $X = Y$ reveals how the distributions differ \cite{Loy02042016, wang1998using}.
 \begin{enumerate}
\item  \textbf{Q-Q Plots and Skewed Distributions}: the distribution is left-skewed when the Q-Q plot would be concave downward; the distribution is right-skewed , when the data show a U-shaped or "humped" pattern; and  a Q-Q Plots of any symmetric distribution is typically symmetric and linear in the center of the data.
\item \textbf{Q-Q Plots and Short-Tailed Distributions}: Short-tailed distributions may show an $S$-shaped curve, but the more specific characteristic is the deviation from the straight line in the opposite direction at the tails compared to long tails ( above the line in the lower tail and below the line in the upper tail). 
\item  \textbf{Q-Q Plots and Long-Tailed Distributions\cite{thode2002testing}:} Long-tailed distributions typically show deviations from the straight line at both ends of the Q-Q plot, where  the lower tail turns downward and the upper tail curves upward.
\item  \textbf{$S$-shaped Curves in Q-Q Plots} can indicate several things, including: Heavier tails than the theoretical distribution, Light tails and systematic differences between the distributions being compared. 
 \end{enumerate}
  \end{remark}
 
 \section{Conclusion}
\noindent
This study presents a comparative analysis of the tail behavior in the daily return distributions of Bitcoin and Ethereum using the seven-parameter Generalized Tempered Stable (GTS) distribution. Through rigorous statistical modeling and quantile–quantile (Q–Q) plot analysis, we demonstrate that both assets exhibit significant deviations from normality, characterized by heavy-tailed distributions.
Our results indicate that while Bitcoin returns already display pronounced tail risk, Ethereum's return distribution exhibits even heavier tails and greater process intensity. This suggests a higher likelihood of extreme return events in Ethereum, highlighting its high risk profile relative to Bitcoin. 
The findings have important implications for financial practitioners, particularly in areas such as risk management, algorithmic trading, and derivative pricing, where accurately modeling tail risk is critical. Additionally, the superior fit of the GTS distribution over conventional models highlights its practical value in capturing the complex dynamics of cryptocurrency returns.

 \bibliographystyle{IEEEtran}
\bibliography{fmodelArxiv}

\end{document}